\documentclass[fleqn,usenatbib]{mnras}
\usepackage{graphicx}
\usepackage{amsmath,amssymb}
\usepackage{lastpage}
\usepackage[all]{hypcap}
\usepackage[T1]{fontenc}
\usepackage{float}
\usepackage{subfigure}
\usepackage{afterpage}
\usepackage[usenames]{color}
\usepackage{mathrsfs}
\usepackage{upgreek}
\usepackage{hyperref}
\usepackage{color}
\usepackage[dvipsnames]{xcolor}
\usepackage{xspace}
\usepackage{physics}

\newcommand{\kpch}{kpc$/h$\xspace}
\newcommand{\Mpch}{Mpc$/h$\xspace}
\newcommand{\Gpch}{Gpc$/h$\xspace}
\newcommand{\Msunh}{M$_\odot/h$\xspace}
\newcommand{\cb}{\texttt{CSiBORG}\xspace}
\newcommand{\qj}{\texttt{Quijote}\xspace}
\newcommand{\phew}{\texttt{PHEW}\xspace}
\newcommand{\fof}{\texttt{FOF}\xspace}
\newcommand{\hop}{\texttt{HOP}\xspace}
\newcommand{\ramses}{\texttt{RAMSES}\xspace}
\newcommand{\borg}{\texttt{BORG}\xspace}
\renewcommand{\order}[1]{\mathcal{O}(#1)}
\newcommand{\warren}{W06\xspace}

\title{The effect of local universe constraints on halo abundance and clustering}
\author[M.~L.~Hutt et al.]{
Maxwell~L.~Hutt$^{1,2}$\thanks{E-mail: m.hutt22@imperial.ac.uk}, Harry~Desmond$^{1,3,4}$\thanks{E-mail: harry.desmond@port.ac.uk}, Julien Devriendt$^1$ and Adrianne Slyz$^1$
\\
$^{1}$Astrophysics, University of Oxford, Denys Wilkinson Building, Keble Road, Oxford, OX1 3RH, UK\\
$^{2}$The Blackett Laboratory, Imperial College London, Prince Consort Road, London, SW7 2AZ, UK\\
$^{3}$McWilliams Center for Cosmology, Department of Physics, Carnegie Mellon University, 5000 Forbes Ave, Pittsburgh, PA 15213\\
$^{4}$Institute of Cosmology \& Gravitation, University of Portsmouth, Dennis Sciama Building, Portsmouth, PO1 3FX, UK
}

\pubyear{2022}

\begin{document}
\label{FirstPage}
\pagerange{\pageref{FirstPage}--\pageref{LastPage}}
\maketitle

\begin{abstract}
Cosmological $N$-body simulations of the dark matter component of the universe typically use initial conditions with a fixed power spectrum and random phases of the density field, leading to structure consistent with the local distribution of galaxies only in a statistical sense. It is, however, possible to infer the initial phases which lead to the configuration of galaxies and clusters that we see around us. We analyse the \cb suite of 101 simulations, formed by constraining the density field within $155$ Mpc$/h$ with dark matter particle mass $4.38\times10^9$ M$_\odot$, to quantify the degree to which constraints imposed on 2.65 \Mpch scales reduce variance in the halo mass function and halo--halo cross-correlation function on a range of scales. This is achieved by contrasting \cb with a subset of the unconstrained \qj simulations and expectations for the $\Lambda$CDM average. Using the \fof, \phew and \hop halofinders, we show that the \cb suite beats cosmic variance at large mass scales ($\gtrsim10^{14}$ \Msunh), which are most strongly constrained by the initial conditions, and exhibits a significant halo--halo cross-correlation out to $\sim30$ \Mpch. Moreover, the effect of the constraints percolates down to lower mass objects and to scales below those on which they are imposed. Finally, we develop an algorithm to ``twin'' halos between realisations and show that approximately 50\% of halos with mass greater than $10^{15}$ \Msunh can be identified in all realisations of the \cb suite. We make the \cb halo catalogues publicly available for future applications requiring knowledge of the local halo field.
\end{abstract}


\section{Introduction}
\label{sec:intro}
Undetectable except by its gravitational effect, dark matter (DM) is thought to be the dominant matter component of the universe. Maps of large volumes of the universe find galaxies to be arranged in a web-like structure of clusters, filaments and voids, tracing the distribution of the underlying dark matter density field. Understanding the evolution of this density field and its connection to visible tracers are currently key tasks in astrophysics.

Simulations of DM with initial conditions (ICs) based on a fixed power spectrum and random phases of the primordial matter field have been well studied for decades \citep{Press74, Davis85, Lacey94, Eisenstein98, Tinker08}. Only more recently however have attempts been made to constrain the phases as well (i.e. the full 3D initial density field) so as to produce structures with positions matching those observed in the local universe. This subfield originated with \citet{Hoffman_Ribak}, who developed a method for constructing constrained realisations of Gaussian random fields. Early applications of this algorithm developed galaxy count \citep{Kolatt,Bistolas} and peculiar velocity observations \citep{Weygaert,Klypin,Kravtsov} as means of deriving the constraints. Subsequent refinements increased the resolution of the simulations and the precision of the gravity model, improved the modelling used to generate the ICs (e.g. the galaxy bias), sampled the ICs rather than generating them in a maximum-likelihood sense, and implemented additional physics. Galaxy groups have also been used for deriving the constraints \citep{ELUCID_I}. A focus of constrained simulations has been the properties and formation history of the Local Group \citep{CLUES, CLUES_Sorce, Sibelius, Sibelius-dark}, although they are finding use in an increasingly wide range of science applications.

It was realised already a decade ago \citep{Jasche_first_BORG,Jasche_Wandelt_2,Jasche_Wandelt_1} that deriving the ICs of the local universe is best viewed as a Bayesian inference problem and solved by forward modelling. The method for doing this---which expanded to include full models for galaxy bias, redshift space distortions and observational selection effects---became known as the \emph{Bayesian Origin Reconstruction from Galaxies} (\borg) algorithm, which has been applied to the 2M++ \citep{2M++,BORG_2M++,BORG_2M++_PM} and SDSS-III/BOSS \citep{BORG_SDSS} galaxy fields. This uses an efficient Hamiltonian Markov Chain Monte Carlo algorithm to sample the ICs at $z=1000$ as well as the nuisance parameters. Although other constrained simulations have been run using ICs derived with similar techniques (e.g. \citealt{ELUCID_I,ELUCID_II,ELUCID_III}), there has not before now been a statistical ensemble of such simulations that fully samples the IC posteriors. This is needed to propagate the uncertainties in the IC inference into the low-redshift density field at high resolution, and hence quantify the extent to which our present knowledge of the local universe's ICs permits us to localise structures of various masses on various scales and reduce cosmic variance in their $N$-point statistics. Such an ensemble is also necessary to marginalise over the IC uncertainties in post-processing applications.

To fill this gap, we analyse \cb (\textit{Constrained Simulations in BORG}), a suite of constrained simulations with ICs derived from the 2M++ particle mesh \borg reconstruction \citep{BORG_2M++_PM}. The 101 realisations of the suite span the converged part of the \borg posterior, allowing us to investigate the effect of the uncertainty it introduces. \cb has been used previously to test galileon modified gravity in \citet{Bartlett21} and to identify local voids-as-antihalos in \citet{Desmond21}. 

Our focus here is on quantifying the effects of the constraints on the variance in halo abundance and clustering (i.e. their uncertainty), as a function of mass and scale, while accounting fully for the \texttt{BORG} IC uncertainties. This is important to understand the exact manner in which constraints on one scale affect structure on others, an important issue for all constrained simulations with ICs specified on a certain scale or set of scales. While the constraints will not affect the finite-volume uncertainty in halo statistics (for fixed box size), they will reduce shot noise on all scales because pinning down larger-scale modes will induce correlations on smaller scales. We assess this by comparing the \cb statistics with those of \qj \citep{quijote}, an unconstrained suite with otherwise similar properties. In particular we address the following questions:
\begin{itemize}
    \item What is the impact of the \texttt{BORG} constraints on the variance in the halo mass function (HMF) between realisations of the local universe?
    \item How do the constraints propagate into the halo--halo cross-correlation function on a range of scales?
    \item Which halos should be considered ``the same'' between the different local-universe realisations, and how does the fraction of matched halos depend on the properties of the halos?
    \item How do these results depend on the halofinder used (Parallel HiErarchical Watershed (\phew, \citealt{PHEW}), friends-of-friends (\fof, \citealt{Davis85}), or \hop \citep{Eisenstein98}?
\end{itemize}

This paper is structured as follows. In Sec.~\ref{sec:simulations} we describe the \cb and \qj simulation suites. In Sec.~\ref{sec:method} we present the halo statistics that we investigate, describe the halofinders used and document our method for matching halos between realisations. In Sec.~\ref{sec:results} we compare the variance in the HMF and halo--halo cross-correlation function of \cb to those of \qj and analytic expectations, and show the results of our matching procedure. In Sec.~\ref{sec:disc_conc} we discuss systematic uncertainties and summarise our results.

\section{Simulated data: \cb \& \qj}
\label{sec:simulations}

The \cb suite comprises 101 realisations of a 677.7 \Mpch box centred on the MW with ICs inferred from the 2M++ survey \citep{BORG_2M++}, spanning the \texttt{BORG} posterior. Within this simulation volume, discretised onto a $256^3$ grid, the \texttt{BORG} algorithm uses a particle-mesh gravity solver \citep{BORG_2M++_PM} to evolve the DM particle field from an initial redshift of $z=1000$ to the present day. This constrains the ICs on a grid of cell length $2.65$ \Mpch.

The density field is well constrained in the volume in which the 2M++ survey has high completeness, a spherical volume with radius $\sim155$ \Mpch around the MW. In this sub-volume the ICs, propagated linearly to $z=69$, are augmented with white noise on a $2048^3$ grid, giving a spatial resolution of $0.33$ \Mpch and a minimum DM particle mass of $3.09\times10^9$ \Msunh. Surrounding this zoom-in region, there is a buffer region in which the DM particle mass increases for distances further from the MW. This buffer extends approximately 10 \Mpch in order to smoothly connect the high resolution central region to the outer region with the same grid resolution as that of \texttt{BORG}. The adaptive mesh refinement code \ramses \citep{ramses} is then used to evolve the DM density field to $z=0$. Only the central augmented region is refined (reaching level 18 by $z=0$, a spatial resolution of 2.6 \kpch), so it is only there that we consider halos. The cosmology of \texttt{BORG} \citep{BORG_2M++_PM}, and hence the simulations, is primarily taken from the \citet{Planck2014} best fit results including WMAP polarisation, high multipole moment, and baryonic acoustic oscillation (BAO) data, with the exception of $H_0$ which is taken from the 5-year WMAP results combined with Type Ia supernovae and BAO data \citep{WMAP2009}: $T_\text{CMB} = 2.728$ K, $\Omega_m = 0.307$, $\Omega_\Lambda = 0.693$, $\Omega_b = 0.04825$, $H_0 = 70.5$ km s$^{-1}$ Mpc$^{-1}$, $\sigma_8 = 0.8288$ and $n = 0.9611$.

To quantify the effects of the constraints we compare with unconstrained simulations. Ideally, one would run another suite of $\sim100$ simulations identical to \cb except with random ICs. As this is computationally very expensive (the full \cb suite required $\sim$3M CPU-hours), we instead investigate the publicly available \qj simulations. Each of these is an N-body simulation evolved forward in time from unconstrained ICs at $z=127$ to $z=0$, using the \texttt{GADGET-III} code \citep{GADGET3}. The volume of each \qj box is $1\: (\text{Gpc}/h)^3$. We use the 100 realisations with $1024^3$ resolution, giving a particle mass of $8.72\times10^{10}$ \Msunh, and their ``fiducial cosmology'' ($\Omega_m = 0.3175 $, $\Omega_\Lambda = 0.6825$, $\Omega_b = 0.049$, $H_0 = 67.11$ km s$^{-1}$ Mpc$^{-1}$, $\sigma_8 = 0.834$ and $n = 0.9624$), with the phases of the ICs varying randomly between realisations at fixed power spectrum. This subset is chosen as its resolution is high enough to compare to \cb over a broad mass range. In future, the development of suites of matched constrained and unconstrained simulations would eliminate any potential biases due to differences between the simulations other than their ICs.

To isolate the effect of the IC constraints, we must mitigate the impact of the different cosmological parameters used by \cb and \qj. We do this by running two simulations with unconstrained ICs drawn from a Gaussian density field at $z=69$ with the same resolution, box size and refined sub-volume as the \cb simulations. One of these has the cosmology of \cb and the other that of \qj. Again, we consider only halos located in the central 155 \Mpch ``zoom-in'' sphere. These boxes link the two simulation suites since they have identical resolution and ICs and differ only in their cosmological parameters. 

\section{Method}
\label{sec:method}
In this section we describe the methods used for halofinding, the statistics that we compare between the simulations described above and the procedure for matching halos between realisations.

\subsection{Halofinding}
\label{sec:halofinding}

We run three halofinders on the \cb boxes and the two additional unconstrained simulations: \phew, \fof and \hop. For \qj we use only the \fof algorithm that has already been run on it by \citeauthor{quijote}.

The watershed halofinder \phew was run on-the-fly with \ramses for the \cb simulations. This algorithm splits the density field into basins by locating watersheds (boundaries between regions which share a common local minimum density via steepest descent). These basins are sequentially merged according to user-defined thresholds on the density of saddle points between basins. We use the standard threshold value from \citeauthor{PHEW} of $200\rho_c$ to merge sub-halos into halos. 

For \fof we adopt a linking length parameter of $b=0.2$. This matches the choice made for the \qj simulations by \citeauthor{quijote}. \fof links together particles within a given ``linking distance'' of each other, defined as $b$ times the mean inter-particle separation. It is known that \fof tends to connect extraneous particles to halos along nearby filaments of the density field \citep{Eisenstein98}, creating more massive structures than other algorithms. This is particularly true for halos which are merging near $z=0$ as \fof is more likely to cluster them into a single object than other algorithms, an effect similar to the ``bridged" halos discussed by \citet{Lukic09}. \citet{Warren06} also find that \fof halos are biased towards larger masses for halos with low particle number and propose a correction to the particle number of the form $N_\text{corrected} = N_0(1-N_0^{-0.6})$ for this effect, with $N_0$ the original number of DM particles in the halo. Marginally decreasing the particle number in this way was found to eliminate the effect of resolution on halo mass. \citeauthor{Lukic09} have shown that corrections of this sort must depend on halo concentration as well as particle number. For our purpose of quantifying the variance between realisations, however, this level of precision of the HMF is not necessary and we do not use such corrections.

Finally, we run \hop. This algorithm first generates an estimate of the local density at each particle position using a kernel over a number $N_\text{dens}$ of neighbouring particles. Then it associates to each particle the highest density particle from its nearest $N_\text{hop}-1$ neighbours. This association is repeated until a local maximum of the density field is reached. All particles above a certain density threshold $\rho_t$ which lead to the same local maximum are then grouped together. We run \hop with parameters $N_\text{hop} = N_\text{dens} = 20$ and $\rho_t=80$, which produces similar results to \fof with $b=0.2$ (see \citealt{Tweed09} for a summary of these parameters in the \texttt{AdaptaHOP} \citealt{Aubert04} algorithm, which reduces to \hop when substructure is neglected).

In order to reduce resolution errors in halo properties we require that halos found by any halofinder must contain at least 100 particles. This corresponds to a minimum halo mass of $3.09\times10^{11}$ \Msunh for \cb and $8.72\times10^{12}$ \Msunh for \qj. However, the lower resolution of the \qj simulations means that this constraint would severely limit the mass range across which halos are found compared to \cb. Hence, we include halos from the \qj suite with a minimum of 20 DM particles when it is desirable to extend the comparison with \cb to lower masses.

\subsection{The halo mass function and halo--halo cross-correlation function}
\label{sec:stats}

We compare halo distributions via two main summary statistics: the HMF, defined by
\begin{equation}
    n(M) \equiv \dv{n}{\log M}
\end{equation}
where $\log$ is the base-10 logarithm throughout, and the two-point halo--halo cross-correlation function between pairs of simulation boxes. Denoting the halo density field by $\rho(\vb{x})$ and its mean by $\bar{\rho}$, we measure the cross-correlation
\begin{equation}
    \xi_{ij}^{(hh)}(\vb{r}) = \frac{1}{V}\int\dd[3]{\vb{x}} \delta_i(\vb{x}) \delta_j(\vb{x}-\vb{r})
\end{equation}
where the subscripts label the pair of simulation boxes being compared and
\begin{equation}
    \delta_i(\vb{x}) \equiv \frac{\rho_i(\vb{x})}{\bar{\rho}_i} - 1
\end{equation}
is the halo overdensity field. Assuming an isotropic universe, $\xi(\vb{r}) = \xi(r)$ and we can estimate the cross-correlation by counting pairs of halos separated by distances which fall into a given set of radial distance bins and adopt the Landy-Szalay estimator \citep{Landy93}
\begin{equation}
    \hat{\xi}_{ij}^{(hh)} (r) = \frac{D_iD_j-D_iR-D_jR+RR}{RR}
\end{equation}
where $D_i$ represents the catalogue of data points (halos) in simulation $i$ and $R$ is a catalogue of random points. That is, $D_iD_j(r)$ is calculated as the number of pairs of data points with separation, $r$, falling in a given distance bin, and similarly for the other terms. We use 100 times as many random points as data points so that the uncertainty in the estimator itself (which is at the Poisson level) is negligible compared to the variance between realisations. We increase the statistical significance by averaging the cross-correlation function over all pairs of realisations in a given suite (\cb or \qj), and denote the averaged quantity simply as $\ev{\xi(r)}$.

The HMF is calculated in each box using 20 bins spaced evenly in $\log(M/(M_{\odot}/h))$. Using a larger number of bins has no impact on the results at the low-mass end but decreases the visibility of trends at the high-mass end where shot noise variance is large as there are fewer halos. For each suite we then find the mean and standard deviation of the HMF, as well as the correlation coefficients between the halo counts in each mass bin (normalised covariance), over all the realisations.

The variance of statistics calculated from $N$-body simulations, often called ``cosmic variance'', has several origins (see \citealt{Colombi99, Colombi94, Moster11} and references therein). The most important are (a) shot noise due to the finite number of objects in the simulation, (b) finite-volume uncertainty due to underlying density modes on scales larger than the simulation volume, and (c) edge effect uncertainty, which is related to the geometry of the survey in that objects near its edge carry less weight than those far from the edge \citep{Szapudi96}. The latter two sources of variance are negligible compared to shot noise at scales approaching homogeneity ($\sim500$ \Mpch; \citealt{Schneider16}). The ICs of \cb will not influence finite-volume or edge-effect uncertainties over the entire simulation box, but the density field constraints will reduce shot noise. By reducing variance in the density modes at and above the IC scale ($\sim 2.65$ \Mpch), the constraints do however induce correlations on smaller scales which effectively reduce finite-volume variance for them too. We will quantify the extent to which constraints determine structure both within and beyond the range of scales on which they are directly imposed.

Since the full \qj volume of 1 \Gpch is well beyond the homogeneity scale, those boxes are not subject to such large finite-volume uncertainties and are therefore dominated by shot noise. However, given that the refined 155 \Mpch region of the \cb simulations is far below it, we cannot use the same method to isolate the shot noise contribution to the \cb HMF variance. We proceed by using the full \qj volume to verify that the shot noise uncertainty is Poissonian, allowing us to infer the shot noise contribution within the refined region.

The logic is as follows. First, we measure the standard deviation $\sigma_{155}$ of the HMF of both the \cb and \qj simulations within $V_{155} = 4\pi \,(155\text{ Mpc}/h)^3/3$, such that finite-volume and edge-effect uncertainties are held constant. Second, we measure the \qj HMF in the full $V_{1000}=(1$ Gpc$/h)^3$ volume for all the \qj simulations and calculate its standard deviation $\sigma_{1000}$. This allows us to confirm that the HMF measured in the smaller volume is indeed consistent with that found over the whole box, and to estimate the isolated shot noise uncertainty. We can then verify that the shot noise uncertainty is Poissonian by computing the expected Poisson variance in $V_{155}$, namely $(N_\text{halo})^{-1/2}$, and comparing it with the inferred variance from the full \qj volume, $\sigma_{1000}\sqrt{V_{155}/V_{1000}}$. $N_\text{halo}$ is the average number of halos in the 155 \Mpch region over all realisations. If these two measurements of the shot noise are consistent then we can be confident that the shot noise error in $V_{155}$ is given by $(N_\text{halo})^{-1/2}$ and we can assume this for \cb also. Thus we isolate the shot noise contribution to both the \qj and \cb HMF variances and can identify whether the constraints on \cb decrease the variance below the Poissonian expectation.

We compare the HMF to a standard function from the literature, \citet[hereafter \warren]{Warren06}. We choose this particular analytic form because it was originally fitted to the results of a \fof halofinder on a wide range of unconstrained $N$-body simulations. Other works \citep{Despali15} have shown that improved parametrisations and halo definitions lead to HMFs which are $\sim30\%$ discrepant from \warren at large halo masses. We do not, therefore, expect this curve to be fully compatible with our results, but it does allow us to compare the \cb and \qj HMFs to a standard often seen in the literature.

\subsection{Matching halos between realisations}
\label{sec:halo_matching}

We now describe a method for identifying halos that are sufficiently similar in mass and location between all realisations of a simulation suite to be called the ``same'' object. The halos that are then present in \emph{all} realisations are particularly robust and should potentially be prioritised in post-processing applications. For this, we use the \fof halo catalogues of \cb and consider halos with masses greater than $10^{13}$ \Msunh to enable comparison with \qj over the same mass range. The procedure is as follows:
\begin{enumerate}
    \item Select a reference simulation and record the positions of all halos in it.
    \item Select a halo in the reference simulation and calculate its size according to $R_{200} = (3M/(4\pi\cdot200 \rho_\text{crit}))^{1/3}$. 
    \item In all other simulations, locate all halos within a ``search radius'' around the position of the reference halo selected in the previous step and select the one most similar in mass.
    If there are any simulations in which \emph{no} halos are located within the search radius around the reference halo position, discard that halo as it is not consistently present in all simulations. This results in a list of the halos which are ``matched'' to the reference halo in all the other simulations.
    \item Calculate the variance in the mass and position (i.e. distance from the reference halo position) of the matched halos across all simulation boxes. This gives a measure of the success of the matching for this particular halo in the reference simulation.
    \item Repeat from step (ii) until all the halos in the reference simulation have been processed.
    \item Repeat from step (i) until all simulation boxes have been used as the reference simulation.
\end{enumerate}

There are several possible ways to select the ``search radius'' in step (iii). The simplest---a fixed distance in Mpc---would bias the fraction of matched halos high at low mass relative to high mass because more massive, larger halos would be expected to fluctuate more in position than lower mass ones. The next simplest is to scale the search radius with the halo radius, i.e. $NR_{200}$ with constant $N$; this is our fiducial scheme with $N=5$. Although this avoids the previous bias, it may miss low mass halos that could be matched with a more realistically lenient scheme. This is because the deviation in position of halos between the realisations is related to the halos' peculiar velocities. The velocity power spectrum peaks at larger radius than the matter power spectrum so that halos peculiar velocities tend to be larger than their virial velocities \citep{Sheth01,Hahn15}. Thus one might expect lower mass halos to deviate by more than $5R_{200}$ between the realisations, which would prefer a different function for the search radius. Our choice is conservative in the sense that it only associates low-mass halos that are closer together than the expectation from the velocity field, and a custom matching procedure to suit the needs of the user may readily be applied to our public halo catalogues.

In our fiducial model, the exact ratio of the virial and search radii, $N$, is somewhat arbitrary but does not qualitatively affect the salient results: the variance in position and mass of the matched halos, the difference in the number of matched halos in \cb and \qj, and the dependence of this number on mass. We implement the procedure for $N=3,5,10$ to verify this. Note that the search radius around a reference halo with mass $10^{13}$ \Msunh is $5R_{200} \approx 1.8$ \Mpch whereas for a very massive $10^{15}$ \Msunh halo, it is approximately 8 \Mpch. Hence, for the smallest halos being considered, we expect far greater variance in the matching since the search radius is below the IC constrained scale. Nevertheless, identifying more successful matches at low masses in \cb than \qj is a strong indicator of the induced correlation on scales below those set by the \cb constraints.

There is, however, a subtle residual bias in the matching procedure towards smaller halos since within a given search radius in any realisation, the probability to find a small halo is far greater than that of a large halo simply because there are more small halos than large ones. Consider the case of a very massive reference halo. When searching for a match in another simulation, it is possible that there is no halo of similar mass within the search radius. In this case, the procedure is more likely to match this reference halo to a small halo which is found there. Hence, the mean mass of the matched halos tends to be slightly lower than the mass of the reference halo. In our application, there are negligibly many ($\order{0.1\%}$) halos where the mean mass of matched halos differs by more than a factor of 5 from the reference halo mass and hence we have not imposed any filters on the matched halos. In general, one may reduce this bias by imposing a threshold variance on the mass of the matched halos, or on the ratio of the mean matched halo mass to the reference halo mass.

We show the result of this procedure in Fig.~\ref{fig:matched_halos} by plotting the variance in the mass and position of all the matched halos as a function of the reference halo mass, with the expectation that larger halos will be more reliably matched between simulations because they are better constrained by the \cb ICs. We also run this procedure on the \qj suite, in order to determine the proportion of halos one would expect to find suitable matches for in the absence of IC constraints. 

\section{Results}
\label{sec:results}

\subsection{\cb \& \qj HMFs}

\begin{figure}
    \centering
    \includegraphics[width=0.5\textwidth]{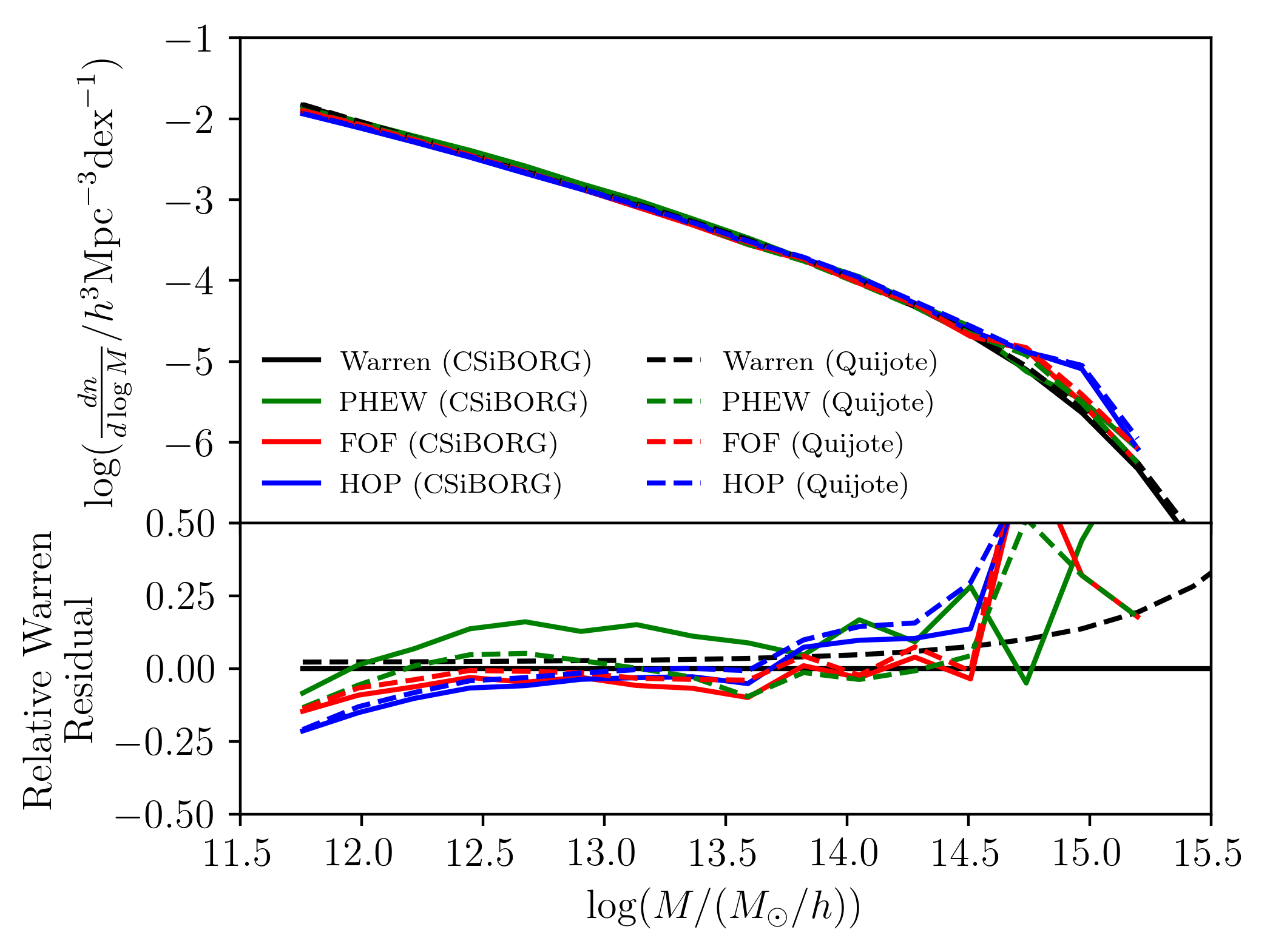}
    \caption{\emph{Upper panel:} HMF of the pair of simulations with identical unconstrained ICs but differing in their cosmological parameters calculated with different halofinders. The solid lines use the \cb cosmology and the dashed lines the \qj cosmology. (Note that none of these boxes are part of the actual \cb or \qj suite; the labels ``\qj'' and ``\cb'' refer merely to the cosmology.) The black curves show the HMF from \warren. \emph{Lower panel:} Relative residual from \warren using the \cb cosmology.}
    \label{fig:HMF_cosmology}
\end{figure}

We begin our investigation of the HMF by quantifying the shift in it due to the different cosmologies of \cb and \qj. Fig.~\ref{fig:HMF_cosmology} shows the HMFs of the two unconstrained simulations performed using the \cb resolution with identical ICs but different cosmologies, as described in Sec.~\ref{sec:simulations}, using the three halofinders. We show also a pair of curves using the HMF from \warren, evaluated for both cosmologies, as a point of comparison.

\begin{figure}
    \centering
    \includegraphics[width=0.48\textwidth]{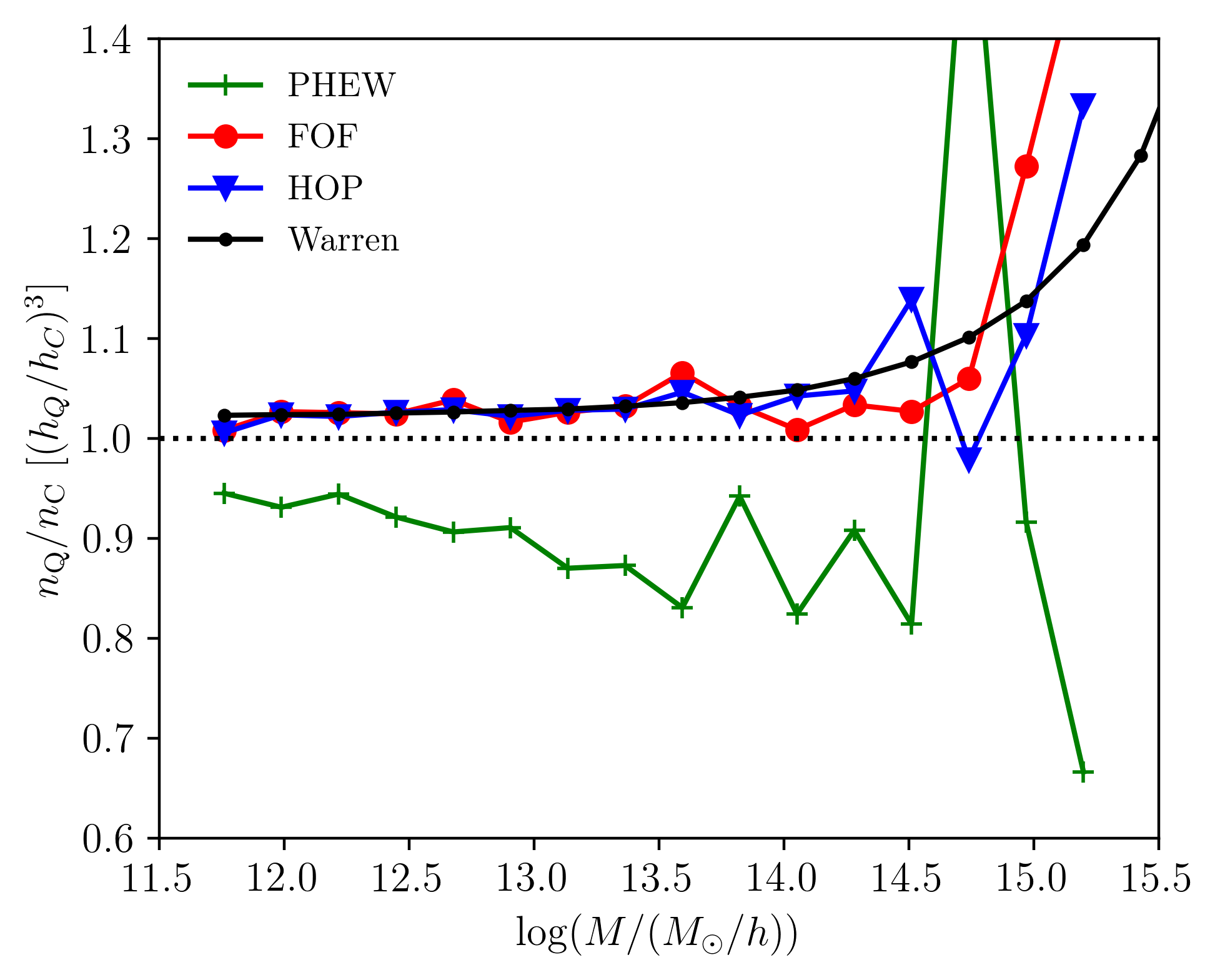}
    \caption{Ratio of HMFs of the \qj and \cb cosmologies shown in Fig.~\ref{fig:HMF_cosmology} for the different halofinders. The black curve is the ratio found using the \warren HMF. In the axis label, the subscripts ``$C$'' and ``$Q$'' refer to the \cb and \qj cosmologies respectively.}
    \label{fig:HMF_cosmology_ratio}
\end{figure}

It is apparent from this plot that, for a fixed cosmology, \phew exhibits different behaviour to \fof and \hop: its HMF is $>10\%$ greater for masses between $10^{12}$ and $10^{14}$ \Msunh and markedly lower at very large masses. \phew also shows fundamentally different behaviour to the other halofinders between the two cosmologies, as shown in Fig.~\ref{fig:HMF_cosmology_ratio}. The \phew HMF is larger in the \cb cosmology than in the \qj one whereas the curves from \warren, \hop, and \fof have the opposite behaviour. Since this work focuses on the variance between different realisations, we do not investigate further this discrepancy between \phew and the other halofinders. In light of this result we do, however, regard the FOF and HOP results as more reliable.
In Fig.~\ref{fig:HMF_cosmology}, we note also a decline in the HMF from all three halofinders at $M\lesssim 10^{12.5}$ \Msunh compared to the fiducial curves. This is due to a variety of numerical effects affecting halos containing fewer than $\sim$ 300 particles, including the starting redshift of the ICs, the order of the Lagrangian Perturbation Theory used to generate these \citep{Hahn, Michaux} and the gravity solver used to run the simulation (e.g. \citealt{Schneider16}).

\begin{figure}
    \centering
    \includegraphics[width=0.48\textwidth]{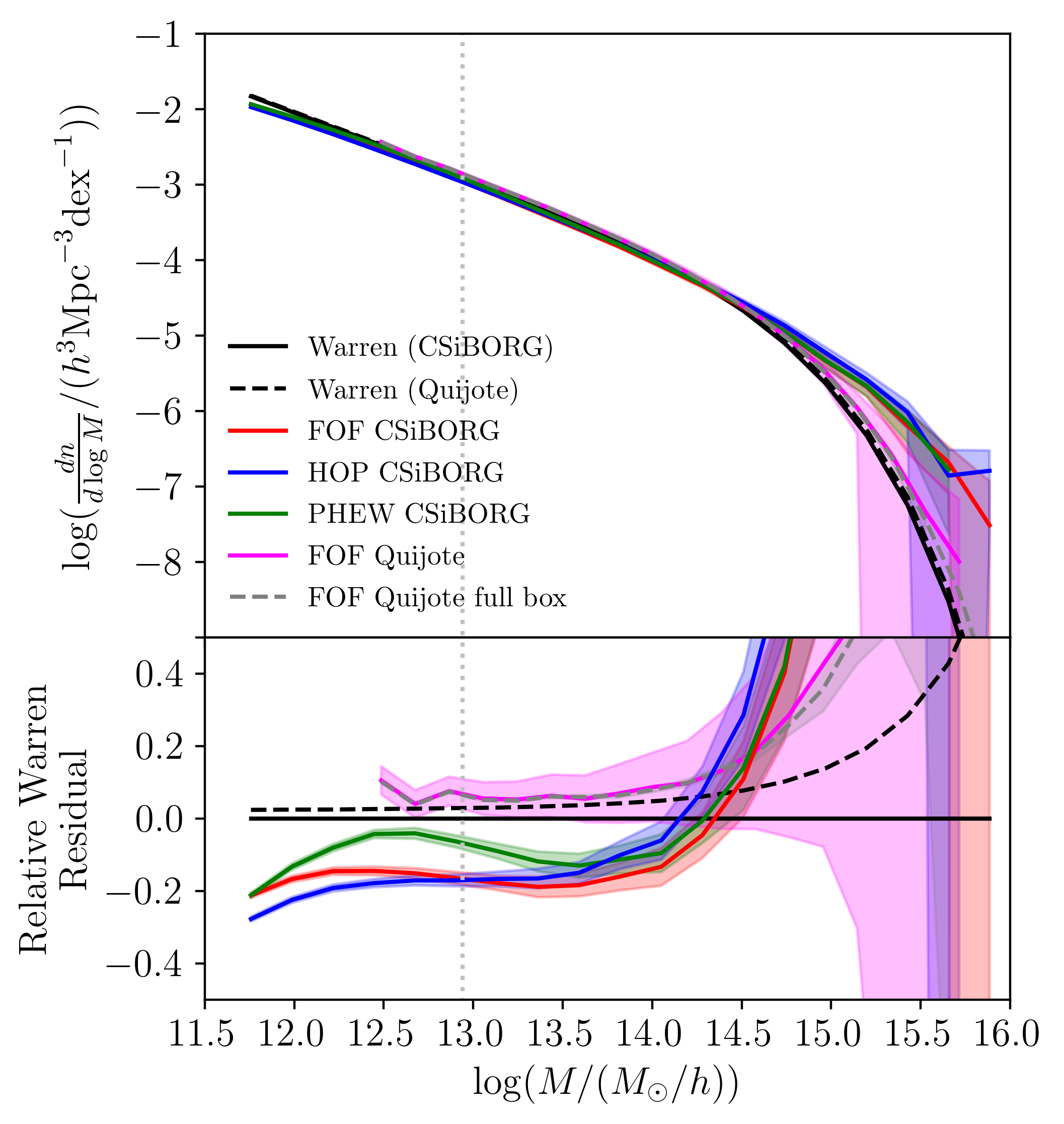}
    \caption{\emph{Upper panel:} Mean HMFs of the \cb suite using all three halofinders and of the \qj suite using the \fof halofinder both in the full (1 Gpc$/h$)$^3$ volume (grey dashed) and the spherical 155 \Mpch volume (pink). The shaded regions around each curve show $\pm1\sigma$ standard deviation estimated across all the simulations in a suite. The black curve shows the HMF from \warren, calculated in the \cb cosmology (solid line) and \qj cosmology (dashed line). \emph{Lower panel:} Relative residual from \warren using the \cb cosmology. The light grey vertical dotted line indicates the mass at which halos in the \qj suite contain 100 particles.}
    \label{fig:HMF_mean}
\end{figure}

We now show the HMF of the \cb and \qj suites, averaged over all simulation boxes, alongside the result from \warren evaluated using the \cb cosmology (Fig.~\ref{fig:HMF_mean}). The \qj curve is calculated in a 155 \Mpch spherical volume to allow comparison to the \cb ``zoom-in'' region, although we also show the \qj HMF of the full (1 Gpc$/h$)$^3$ volume with a dashed grey line and standard deviation band between realisations. The \qj lines include all halos with $\geq20$ particles to enable comparison to \cb down to lower mass, although we caution that halos with fewer than 100 particles ($M<8.72\times10^{12}$ \Msunh for \qj) have mass uncertainty at the $\sim$30--50 per cent level due to imperfect resolution \citep{Trenti10}.

The \qj HMF begins to exceed the \warren curve corresponding to the same cosmology considerably for $M \gtrsim 10^{14.5}$ \Msunh. This behaviour is well documented \citep{Eisenstein98} and is a result of the greatly varying geometries of \fof halos: the algorithm tends to link particles along filaments of the cosmic web, increasing the halo mass in a way which is not observed for algorithms which produce halos with more consistent geometrical structure. We find that the \cb HMFs overshoot \warren even further in this mass range, however, and also fall considerably below it at lower masses. At low masses, all halofinders produce a HMF $\sim 20\%$ below \warren and at the largest masses probed ($\sim 10^{15.7}$ \Msunh) the \cb HMF is $\sim20$ times ($1.5\sigma$) larger than that of \qj. This discrepancy cannot be accounted for by a 5\% reduction in the value of $\Omega_m$ used to calculate the \warren HMF (which corresponds to the relative underdensity of the \cb volume compared to the cosmic average). The \borg reconstruction of the 2M++ volume therefore implies more massive structures, and fewer low-mass scale structures, than average. While this is not highly statistically significant at high mass due to cosmic variance (nor in the density field itself; \citealt{Sibelius-dark}), it is at lower masses. These results may indicate an issue with the 2M++ \borg ICs (J. Jasche \& G. Lavaux, priv. comm.) which does not, however, appreciably affect the variance between realisations, the focus of our work here.

We also find differences between the results from different halofinders. The three halofinders produce consistent results at large masses whereas at the low mass end there is a $7\sigma$ difference between the HMFs of \fof and \hop. Despite being close to the \fof result at the smallest halo masses, in the intermediate mass range the \phew HMF has a characteristic local maximum in the Warren residual reminiscent of the trend found in the unconstrained simulations in Fig.~\ref{fig:HMF_cosmology}. We also note that at large masses, the \hop HMF rises consistently above those of \fof and \phew (which agree very closely above $10^{14}$ \Msunh). This behaviour is not seen to such an extent in the unconstrained simulations shown in Fig.~\ref{fig:HMF_cosmology} and the halofinder comparison project of \citet{Knebe11} found the opposite behaviour, with the \fof HMF approximately 30\% greater than that of \hop at the bright end. This could indicate that the relative behaviour of the halofinders is a function of the ICs.

\subsection{Variance of the HMF}
\label{sec:variance}

\begin{figure}
    \centering
    \includegraphics[width=0.5\textwidth]{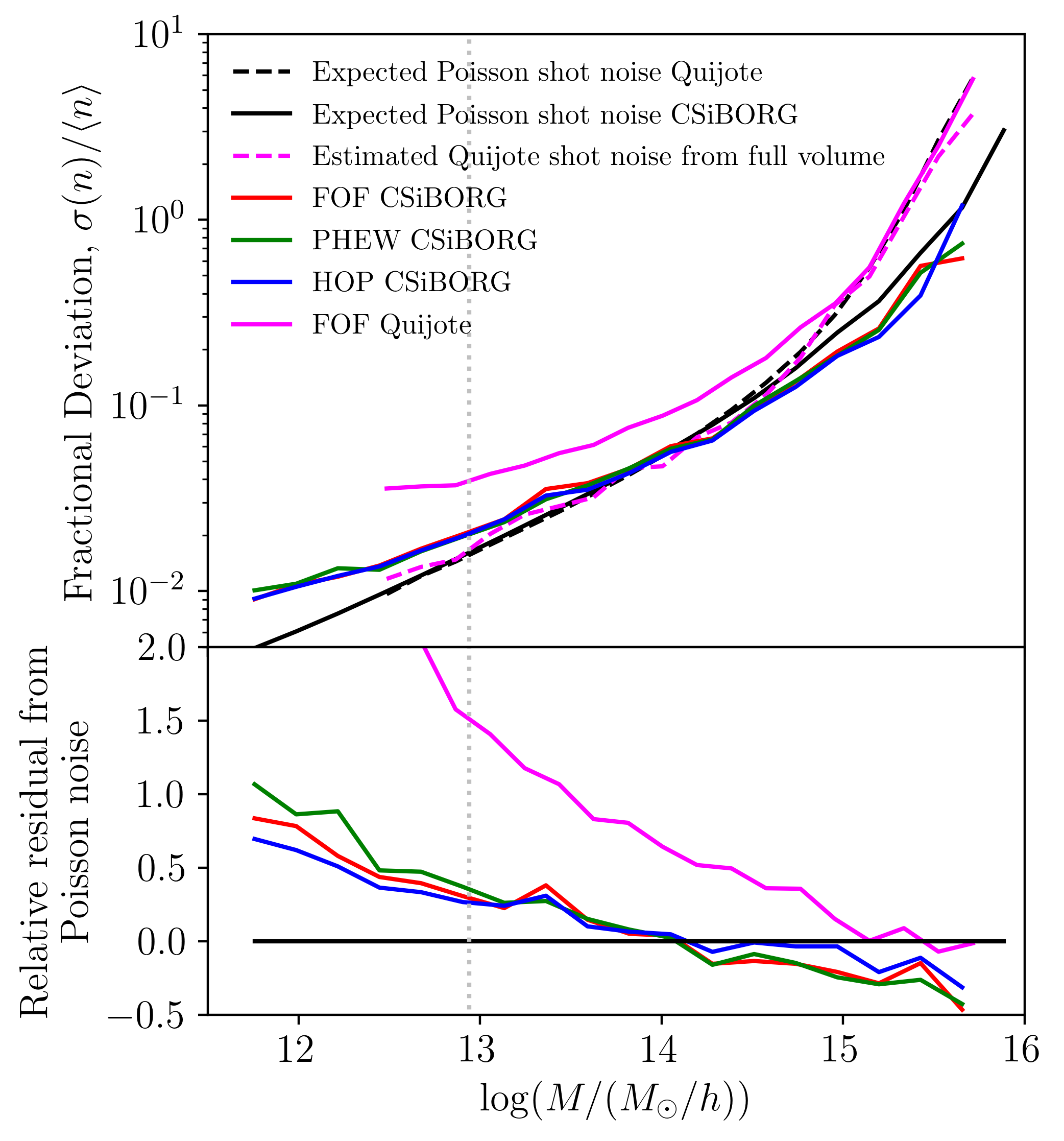}
    \caption{\emph{Upper panel:} Fractional standard deviation of the HMF of the \cb and \qj suites across all simulations in the central 155 \Mpch spherical volume. The dashed pink line is the shot noise contribution to the standard deviation in the 155 \Mpch sphere estimated by normalising the standard deviation measured over the entire \qj volume (see Sec.~\ref{sec:stats}). The expected Poisson shot noise calculated in the 155 \Mpch sphere for the \qj HMF is shown with a dashed black line, and for the \cb \fof HMF with a solid black line. \emph{Lower panel:} Relative residuals of the fractional standard deviation with respect to the Poissonian shot noise expectation for the corresponding HMF. The light grey vertical dotted line indicates the mass at which halos in the \qj suite contain 100 particles.}
    \label{fig:HMF_std}
\end{figure}

We show in Fig.~\ref{fig:HMF_std} the fractional standard deviation of the HMF across all simulation boxes of the \qj and \cb suites (for all halofinders) using the same 155 \Mpch spherical volume for all simulations. We again include all \qj \fof halos down to 20 particles to increase the range of comparison. The fractional deviation of the \qj HMF is approximately a factor of 2 larger than that of \cb across the entire mass range. At the lowest masses, below $1.30\times10^{13}M_\odot$, the fractional deviation of the \qj suite is up to $2.6\times$ larger than that of \cb due to the poorer identification of halos with $<100$ particles.

We compare to the shot noise variance of the \qj HMF in the 155 \Mpch volume, calculated in two separate ways. First we calculate the expected fractional standard deviation due to Poissonian shot noise, $\ev{N_\text{halo}}^{-1/2}$ (using the \qj HMF to estimate $\ev{N_\text{halo}}$) with a dashed black line. Second, as discussed in Sec.~\ref{sec:stats}, we estimate the shot noise contribution in the 155 \Mpch volume by normalising the standard deviation of the HMF calculated using the entire \qj volume across all simulations. This is shown with a dashed pink line. These two methods align closely across all halo masses, which gives confidence that the excess deviation in the \qj HMF calculated in the 155 \Mpch volume above the shot noise contribution is, indeed, due to finite-volume effects, and also that the shot noise error is Poissonian. This also confirms that the dominant source of error at large scales is shot noise since the Poissonian shot noise uncertainty (dashed black) and the measured standard deviation in the 155 \Mpch volume (solid pink) are negligibly different above $10^{15}$ \Msunh. We note in particular that since the \qj simulations are unconstrained, their fractional deviation is never lower than shot noise uncertainty. 

We also show the expected Poissonian shot noise uncertainty in the \cb HMF, $\ev{N_\text{halo}}^{-1/2}$, by using the \fof HMF to estimate $\ev{N_\text{halo}}$, with a solid black line. The lower panel of the figure shows the residual of the measured deviation of the HMF (from a given halofinder) with respect to the corresponding Poissonian shot noise uncertainty. For example, the blue line shows the residual of the measured variance of the \hop HMF with respect to $\ev{N_\text{halo}}^{-1/2}$ where $\ev{N_\text{halo}}$ is estimated using the \hop HMF, and similarly for the other curves. We find a stark difference between the `excess' variance above Poisson shot noise at low masses, which is due to residual finite-volume variance in the relatively small volume used. Since all HMFs are measured in the same region, the finite-volume contribution to the uncertainty is similar between the \qj and \cb simulations (they are not exactly equal as the largest wavelengths of the density field modes differ between the \qj and \cb suite) and therefore the $\sim2$-fold reduction in variance across all masses is a result of the IC constraints in \cb alone.

The true constraining power of the \cb ICs is best evidenced at large masses, $M\gtrsim10^{14}$ \Msunh, where the HMFs measured by all three halofinders have variance below that of the Poissonian expectation. At these halo masses, uncertainty is dominated by shot noise and the \cb variance beats Poisson by as much as 50\%.

\subsection{Covariance of halo masses}
\label{sec:cov}

\begin{figure*}
    \centering
    \includegraphics[width=\textwidth]{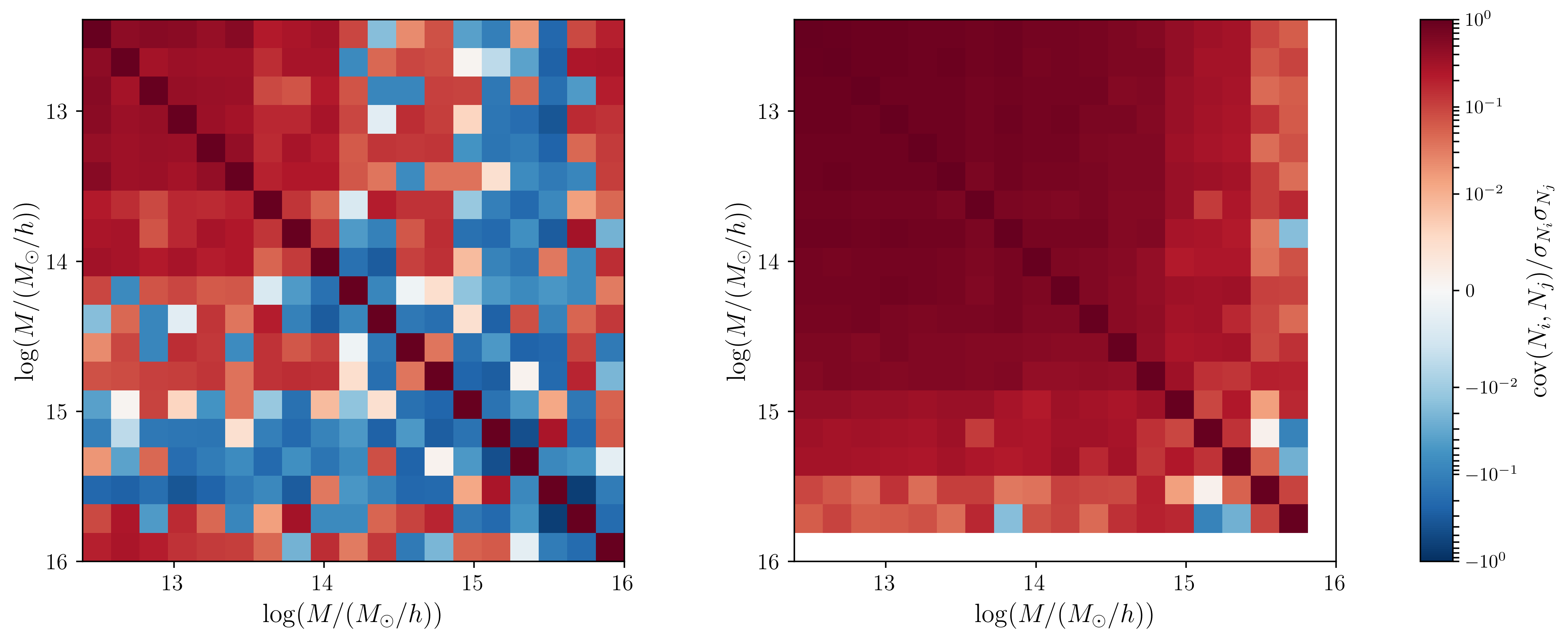}
    \caption{Heatmap of the correlation coefficient matrix of the number of halos in each mass bin for the \fof halofinder in the \cb (left) and \qj (right) suites. The final row and column on the right are left blank as there are no halos in the final mass bin in the \qj suite.}
    \label{fig:correlation_matrix}
\end{figure*}

Fig.~\ref{fig:correlation_matrix} shows the full correlation coefficient matrix of halo counts in \cb and \qj, again going down to 20 particles in the latter. This is the covariance matrix normalised by the standard deviation and hence measures correlations rather than the magnitude of the scatter:
\begin{equation}
    \frac{\text{cov}(N_i, N_j)}{\sigma_{N_i}\sigma_{N_j}} = \frac{\ev{(N_i-\ev{N_i})(N_j-\ev{N_j})}}{\sigma_{N_i}\sigma_{N_j}},
\end{equation}
where $N_i$ is the number of halos in the $i^\text{th}$ mass bin and the averages are taken over all realisations. We use the \fof halofinder for both suites for consistency, although the results for \hop and \phew are extremely similar for \cb.

The dominant effect for the unconstrained \qj suite is that the mass in the central 155 \Mpch region in which we identify halos varies by $\sim8$ per cent between realisations. Realisations with an upward fluctuation of this mass contain more halos at all masses, introducing a global positive covariance. This is particularly significant for less massive halos which respond more uniformly to the fluctuations in the total mass of the volume as they are much more numerous. There are $\order{1}$ halos in the most massive bins in each realisation and halos with masses in excess of $10^{15}$ \Msunh make up less than 10 per cent of the total mass in halos. Hence, even in a realisation with an 8 per cent increase in the total mass of the 155 \Mpch volume, the number of halos in each of these most massive bins is unlikely to vary systematically. This reduces the covariance between the most massive halos in the \qj suite, which, due to statistical fluctuations, is sometimes found to be negative.

Including the IC constraints in the \cb suite, the dominant effect is instead that a halo lost from one mass bin in one realisation is likely to be found in a nearby one in any other realisation. The total mass in halos varies by only $\sim2$ per cent between realisations in \cb, making this effect subdominant to that of halos ``matching'' between realisations with only a small mass difference. This predominantly affects nearby mass bins and massive halos for which the ICs are most constraining; for widely separated bins, or at lower mass, the correlation may again become positive due to the effect of correlated mass fluctuations (although at a weaker level than in \qj due to the lower amplitude of these fluctuations). These results complement those on the variance of the HMF, and illustrate the extent to which halos may be taken to match among the constrained realisations, discussed in more detail in Sec.~\ref{sec:matched_halos}.

\subsection{Cross-correlation function}
\label{sec:cross_correlation}

\begin{figure}
    \centering
    \includegraphics[width=0.5\textwidth]{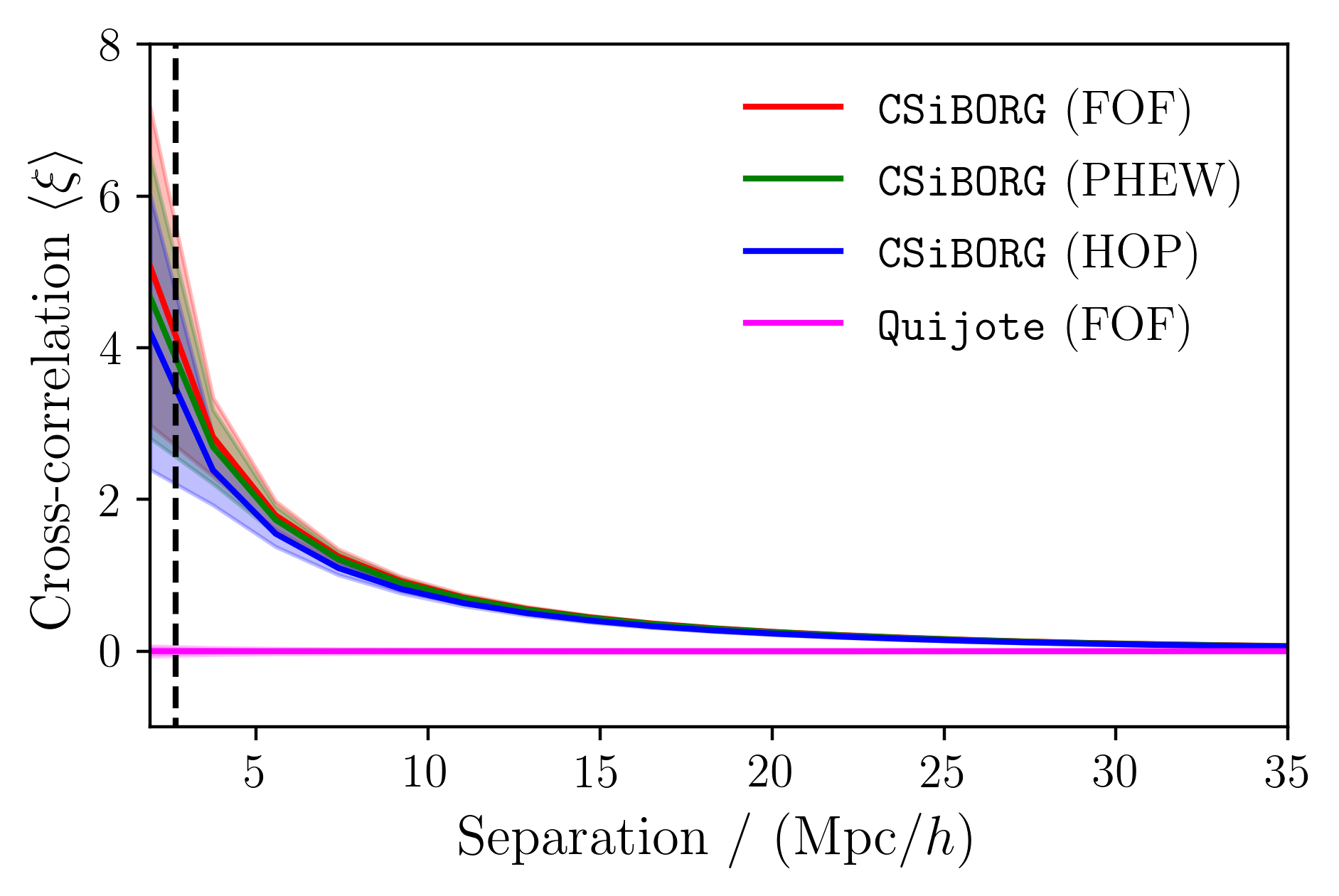}
    \caption{Mean cross-correlation function between all pairs of simulations in the \cb and \qj suites. Only halos with $M>10^{13}$ \Msunh are included. The vertical dashed line shows the IC constrained scale ($2.65$ \Mpch). The error bars show $\pm 1\sigma$ deviation measured in the correlation functions between all pairs of realisations.}
    \label{fig:correlation}
\end{figure}

Fig.~\ref{fig:correlation} shows the two-point halo--halo cross-correlation function between halos of mass in excess of $10^{13}$ \Msunh in the central 155 \Mpch spherical volume, averaged over all pairs of realisations in the the \cb and \qj suites. The solid line shows the average over all pairs of realisations, and the shaded band the standard deviation among them. The different simulation boxes of the \qj suite are fully uncorrelated, as expected. The \cb suite, however, demonstrates strong positive cross-correlation, increasing on smaller scales. This result is salient as the trend continues down to scales below the IC constrained scale ($\sim2.65$ \Mpch). This implies that constraints on large ($>$ \Mpch) scales induce correlation on smaller ones, despite having considerably larger variance at these (formally) unconstrained small scales. This is because smaller-scale modes reside in larger ones, inheriting their constraints to some extent.

\subsection{Matching halos between realisations}
\label{sec:matched_halos}

\begin{figure*}
    \centering
    \includegraphics[width=\textwidth]{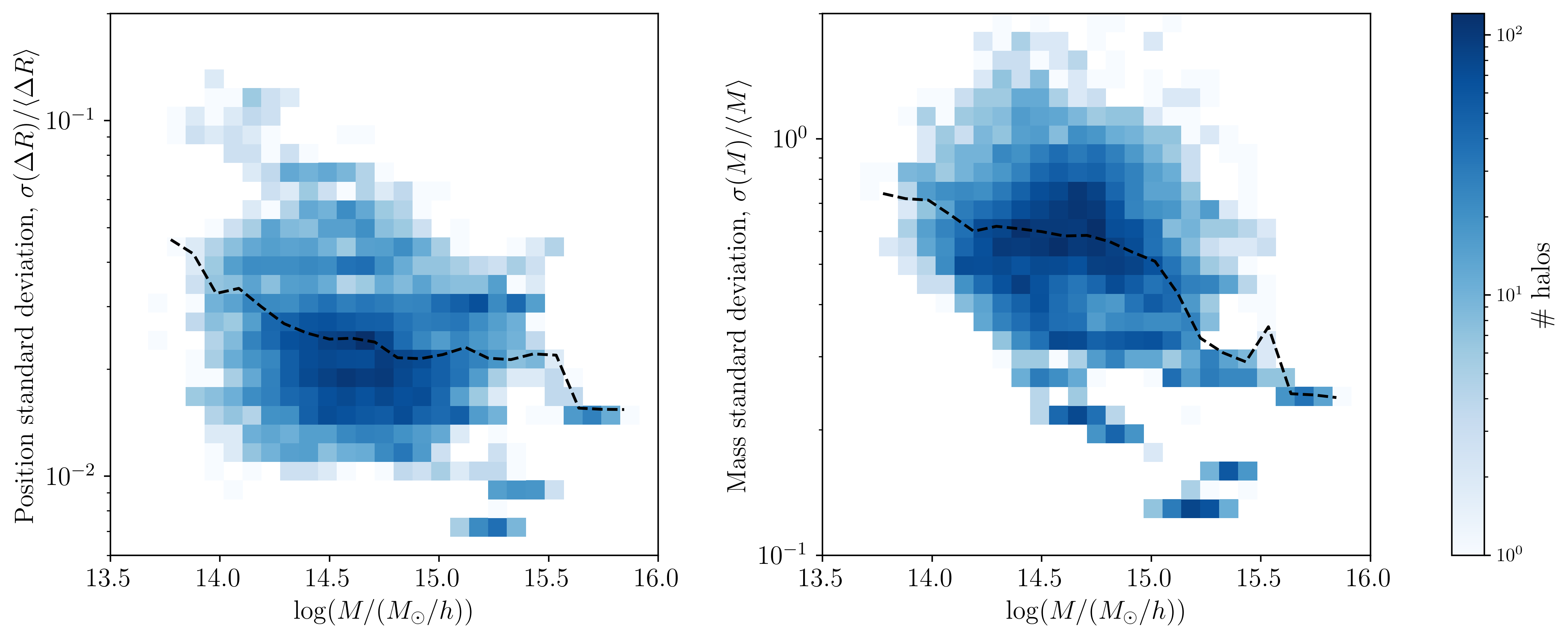}
    \caption{Matched halos distribution for a search radius of $5R_{200}$. \emph{Left panel:} Fractional standard deviation in the position of matched halos with respect to the reference halo position, $\Delta R = \abs{\vb{r}_\text{ref} - \vb{r}}$ where $\vb{r}$ is the matched halo position, as a function of the mass of the reference halo. \emph{Right panel:} Fractional standard deviation in the mass of matched halos as a function of the mass of the reference halo. The dashed black lines trace the mean fractional standard deviation of the matched halos in each mass bin.}
    \label{fig:matched_halos}
\end{figure*}

\begin{figure}
    \centering
    \includegraphics[width=0.5\textwidth]{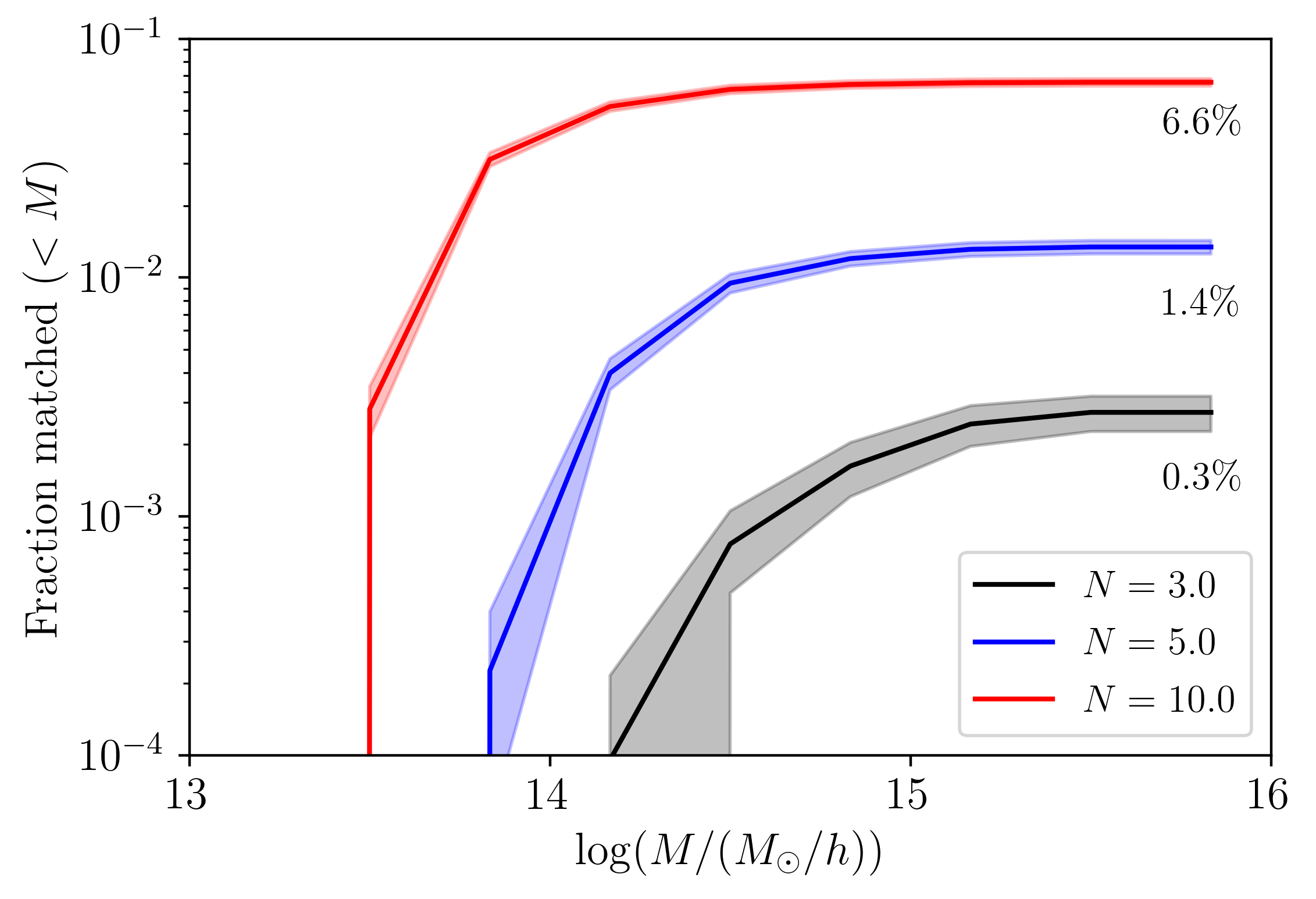}
    \caption{Cumulative fraction of \cb halos matched between all realisations for three different search radii $NR_{200}$ ($N=3,5,10$). The shaded region shows $\pm1\sigma$ standard deviation measured across all reference realisations. The percentages at the far right by each line indicate the total fraction of halos that are matched throughout the entire suite.}
    \label{fig:matching_fraction}
\end{figure}

We show in Fig.~\ref{fig:matched_halos} the fractional standard deviation of the distance between the matched and reference halos, and of the mass of matched halos, when we carry out the halo matching procedure described in Sec.~\ref{sec:halo_matching} on \cb using the \fof halofinder. We find a clear decrease in the relative variance for massive halos compared to lower mass ones, particularly in the matched halo mass. Averaged over all halos of all simulations, for a search radius of $5R_{200}$, approximately 1.4\% of halos are matched throughout all 101 realisations. However, the fraction of halos which are successfully matched varies with the mass of the reference halo, as seen in Fig.~\ref{fig:matching_fraction}. For example, for the intermediate search radius of $5R_{200}$, over 50\% of all halos with masses between $10^{15}$ and $10^{16}$ \Msunh are matched throughout, whereas only a negligible fraction of those with masses $<10^{14}$ \Msunh are present in all realisations. This clearly demonstrates the effectiveness of the constraints for pinning down massive halos.

To make sure our matching procedure depends on the strength of the IC constraints alone, we apply it also to the \qj suite. For $N=5$ we find no matched halos across all simulations at any mass, as we would expect from unconstrained realisations. Hence, the result that 1.4\% of all halos found in a given simulation (and approximately half of the largest halos) are matched successfully throughout the \emph{entire} \cb suite is a statistically significant consequence of the IC constraints. Naturally, unconstrained modes beneath the scale of the constraints lead to structure formation differing between realisations. Our halo matching procedure, as well as the cross-correlation function discussed above, is a strong diagnostic of the strength of the IC inference from \borg, and provides a natural way to quantify the improvement in the strength of the constraints that will be afforded by future inferences. In the public halo catalogues we include a flag that indicates when a halo is matched to a halo in all other \cb realisations when the realisation in question is taken as the reference, using our fiducial matching scheme ($N=5$).

\section{Discussion and Conclusion}
\label{sec:disc_conc}

We analyse the \cb suite of 101 constrained realisations of the 2M++ volume using ICs drawn from the \borg posterior, identifying halos within 155 \Mpch of the MW using three halofinders (\phew, \fof and \hop). We compare the resulting HMFs and halo--halo cross-correlation functions with those of a similar subset of the unconstrained \qj simulations to quantify the extent to which the local universe constraints reduce variance between the realisations. This indicates the reduction in the uncertainty of halo statistics afforded by the \borg algorithm.

By using the entire \qj simulation volume to eliminate finite-volume effects, we isolate the shot noise contribution to the variance in the HMF, and show that the IC constraints of \cb lead to a fractional standard deviation which is $\sim2$ times lower than that of the unconstrained \qj suite over all masses, and up to 50\% lower than Poisson shot noise uncertainty at large masses $\gtrsim10^{15}$ \Msunh. The similarity of halos between \cb realisations is reflected in the covariance matrix of halo masses, which is near-uniformly positive in \qj but typically negative for nearby mass bins in \cb. We also investigate the effect of the constraints on the halo--halo cross-correlation across all pairs of simulation boxes and find positive correlation on all scales, increasing towards short distance. The fact that this trend continues below the IC scale of 2.65 \Mpch implies that large-scale constraints increase consistency in structures on much smaller, formally unconstrained scales. That is, variations in the smaller scale modes of the density field are limited by the constraints on the larger modes which they inhabit. Hence, IC constraints imposed on a given set of scales percolate down to introduce effective constraints, albeit with larger variance, on smaller scales.

We develop a halo matching procedure to identify structures which are particularly well localised throughout the entire \cb simulation suite. The variance in the masses and positions of the matched halos with respect to a given reference halo decreases strongly with increasing mass, indicating that the \cb constraints have greater effect for more massive halos. We demonstrate that the qualitative results of this matching are independent of the details of its implementation, and that our method finds no matches within the unconstrained \qj suite. We publicly release the \cb halo catalogues from the \texttt{PHEW}, \texttt{FOF} and \texttt{HOP} algorithms for analyses requiring knowledge of the local halo field, including its uncertainty given state-of-the-art constraints.

In order to quantify the impact of constrained ICs on the halo population we rely on halofinder algorithms, and our results indicate significant differences between them. In future, the application of 6D algorithms such as \texttt{Rockstar} \citep{rockstar} or \texttt{Velociraptor} \citep{velociraptor} to suites like \cb and \qj would improve halofinding at the massive end because they are less susceptible to erroneously combining halos that are merging near $z=0$.

We anticipate that constrained DM-only and hydrodynamical simulations will be of great value to future studies of the local universe. Future galaxy surveys with high completeness out to scales comparable with the homogeneity scale across large parts of the sky will reduce the finite-volume uncertainties that dominate in small simulation volumes. In parallel, the continued development of algorithms like \borg will integrate the ever-improving observational efforts with the numerical techniques required to pin down the distribution of DM on cosmological scales. Incorporating additional information from, for example, peculiar velocities \citep{Boruah} or cosmic shear \citep{Porqueres} would strengthen the constraints and further reduce variance in the statistics we consider here. An investigation of the effects of ICs imposed on multiple scales would also be illuminating (cf. \citealt{Sibelius} for the Local Group).

\section*{Acknowledgements}
We thank Jens Jasche, Guilhem Lavaux and Stephen Stopyra for helpful discussions. We also acknowledge the authors of the \texttt{Halomod} \citep{Murray21} and \texttt{Corrfunc} \citep{Sinha20} Python libraries, whose codes were used to generate the fiducial HMF curves and the cross-correlation functions respectively.
HD was supported by a Junior Research Fellowship at St John's College, Oxford, a McWilliams Fellowship at Carnegie Mellon University and a Royal Society University Research Fellowship (grant no. 211046). MLH was funded by the President's PhD Scholarships.
This work was done within the \href{https://www.aquila-consortium.org/}{Aquila Consortium}.

This work used the DiRAC Complexity and DiRAC@Durham facilities, operated by the University of Leicester IT Services and Institute for Computational Cosmology, which form part of the STFC DiRAC HPC Facility (www.dirac.ac.uk). This equipment is funded by BIS National E-Infrastructure capital grants ST/K000373/1, ST/P002293/1, ST/R002371/1 and ST/S002502/1, STFC DiRAC Operations grant ST/K0003259/1, and Durham University and STFC operations grant ST/R000832/1. DiRAC is part of the National E-Infrastructure.

This project has received funding from the European Research Council (ERC) under the European Union’s Horizon 2020 research and innovation programme (grant agreement No 693024).

\section*{Data availability}
Our \cb halo catalogues for the \phew, \fof and \hop halofinders are publicly available \href{https://zenodo.org/record/5851241#.Yj89hi-l1mA}{here}. The \qj halo catalogues are available \href{https://quijote-simulations.readthedocs.io/en/latest/}{here}. Other data will be made available upon reasonable request.

\bibliographystyle{mnras}
\bibliography{references}

\bsp
\label{lastpage}
\end{document}